\documentclass[conference,10pt]{IEEEtran}
\IEEEoverridecommandlockouts
\usepackage{cite}
\usepackage{amsmath,amssymb,amsfonts}
\usepackage{algorithmic}
\usepackage{graphicx,physics,array}
\usepackage{textcomp}
\usepackage{xcolor,soul,bm}
\def\BibTeX{{\rm B\kern-.05em{\sc i\kern-.025em b}\kern-.08em
    T\kern-.1667em\lower.7ex\hbox{E}\kern-.125emX}}

\begin{document}

\title{Quantum Random Access Memory For Dummies
}

\author{\IEEEauthorblockN{Koustubh Phalak}
\IEEEauthorblockA{\textit{CSE Department} \\
\textit{Penn State University}\\
PA, USA\\
krp5448@psu.edu}
\and
\IEEEauthorblockN{Avimita Chatterjee}
\IEEEauthorblockA{\textit{CSE Department} \\
\textit{Penn State University}\\
PA, USA\\
amc8313@psu.edu}
\and
\IEEEauthorblockN{Swaroop Ghosh}
\IEEEauthorblockA{\textit{School of EECS} \\
\textit{Penn State University}\\
PA, USA\\
szg212@psu.edu}
}

\maketitle

\begin{abstract}
Quantum Random Access Memory (QRAM) has the potential to revolutionize the area of quantum computing. QRAM uses quantum computing principles to store and modify quantum or classical data efficiently, greatly accelerating a wide range of computer processes. Despite its importance, there is a lack of comprehensive surveys that cover the entire spectrum of QRAM architectures. We fill this gap by providing a comprehensive review of QRAM, emphasizing its significance and viability in existing noisy quantum computers. By drawing comparisons with conventional RAM for ease of understanding, this survey clarifies the fundamental ideas and actions of QRAM. 
\end{abstract}

\section{Introduction} \label{introduction}


Quantum Computing (QC) has progressed rapidly in the past decade. With the advancement in qubit technologies such as superconducting qubits \cite{krantz2019quantum}, trapped ion qubits \cite{bruzewicz2019trapped}, photonic qubits \cite{slussarenko2019photonic}, quantum dots \cite{arakawa2020progress} and diamond nitrogen-vacancy centers \cite{pezzagna2021quantum}, implementation of quantum algorithms on quantum computers has become practically possible. This has also enabled the application of quantum computing in various fields such as machine learning \cite{schuld2015introduction}, finance \cite{herman2022survey}, chemistry \cite{cao2019quantum}, cybersecurity \cite{wallden2019cyber}, and advanced manufacturing \cite{bova2021commercial}. A potential game changer in quantum computing is the augmentation of Quantum Random Access Memory (QRAM) that has shown a potential to provide exponential speedup for algorithms such as, Fourier transform \cite{zhou2017quantum}, discrete logarithm \cite{shor1994algorithms}, and pattern recognition \cite{schutzhold2003pattern, schaller2006quantum, trugenberger2001probabilistic}. QRAM is also a key requirement for important quantum algorithms such as quantum searching of classical database \cite{grover1996fast, brassard2002quantum}, collision finding of hash and claw-free functions \cite{brassard1997quantum} and distinctness of elements in a list \cite{ambainis2007quantum, childs2007weak}. Along with this, a QRAM can also serve as an important memory element to load classical data into the quantum Hilbert space, compared to simpler methods like amplitude, angle, and basis embeddings \cite{schuld2018supervised}.

Existing literature on QRAM fails to summarize key aspects of QRAM and explain them in layman's terms which is the objective of this paper. In \cite{di2020fault}, the authors discuss various QRAMs such as bucket-brigade QRAM, large width small depth QRAM, and small width large depth QRAM but from a fault-tolerance standpoint rather than a fundamental explanatory perspective. 
An overview of the practicality of QRAM in modern NISQ systems is provided in \cite{hann2021practicality} however, it can be esoteric at times to fully comprehend. We provide a simple-to-grasp review of QRAM for readers interested in diving deep into the field of quantum memories. While complex mathematical knowledge of quantum physics is not required, we do assume that the readers know the fundamentals of quantum computing \cite{nielsen2002quantum} such as the ket notation, quantum gates, and quantum circuit notation.  

This paper is organized as follows: In Section II, we provide preliminaries on quantum computing and the workings of classical RAM. In Section III, we delve into the fundamentals of QRAM, answering key questions about its structure, utility, and requirements. Section IV explores the practical implementation of QRAM, while Section V offers an overview of the challenges in implementing QRAM and its future potential. Finally, we conclude in Section VI. 
\section{Preliminaries} \label{prelim}
\subsection{An Overview of Quantum Computing}

\paragraph{\textbf{Qubits}} Qubits, the elementary units of quantum computing, are distinct from classical bits in that they can exist in a superposition of states and represent both 0 and 1 simultaneously. This unique property allows quantum computers to perform multiple computations in parallel, providing the potential for exponential speedup compared to classical computers. In a Hilbert space, a qubit is represented by a two-dimensional vector denoted as $\ket{\psi} = \alpha\ket{0} + \beta\ket{1}$, where $\alpha$ and $\beta$ the coefficients of the basis states of a qubit. These coefficients are constrained by the normalization condition $|\alpha|^2 + |\beta|^2 = 1$, and the probabilities of measuring the state of the qubit in the basis state of $\ket{0}$ or $\ket{1}$ are given by $|\alpha|^2$ or $|\beta|^2$ \cite{nielsen2002quantum}.

\paragraph{\textbf{Quantum Gates}} Quantum gates are the fundamental operations that act on qubits in a quantum circuit, akin to how classical logic gates operate on classical bits. These gates include the Pauli-X, Pauli-Y, Pauli-Z, Hadamard, CNOT, and Toffoli gates. They are often depicted as unitary matrices that act on qubit states, maintaining the quantum properties of the system \cite{rieffel2011quantum}. Quantum gates are created and realized physically utilizing a variety of techniques, including lasers, magnetic fields, and microwave pulses \cite{nielsen2002quantum}. 

\paragraph{\textbf{Quantum Circuit}} Quantum circuits are collections of quantum gates that work together to carry out particular quantum computations. 
Initialization of the qubits is the first step of a quantum circuit. 
Gate operations, such as multi-qubit gates like the CNOT gate and the Toffoli gate, as well as single-qubit gates like the Hadamard and Pauli gates, are used to change the qubits to the required state. Prior to execution, the high-level gates in the circuit, including the Toffoli gate, are disassembled into a native gate set of the quantum hardware (called transpilation). The output of the quantum circuit is then obtained by measuring the qubits using a measurement gate, which converts the quantum state into a classical state \cite{kitaev2002classical}.

\paragraph{\textbf{Quantum Entanglement}} When two or more qubits are correlated in a way that prevents one qubit from being described independently of the other qubits, this phenomenon is known as quantum entanglement. This characteristic is critical for the development of effective quantum algorithms and protocols, including quantum teleportation and superdense coding \cite{bennett1993teleporting}. 
The non-local correlations of entanglement highlight its importance in quantum computing by allowing the execution of tasks that are classically impossible.

\begin{figure}[t]
    \centering
    \includegraphics[width=1\linewidth]{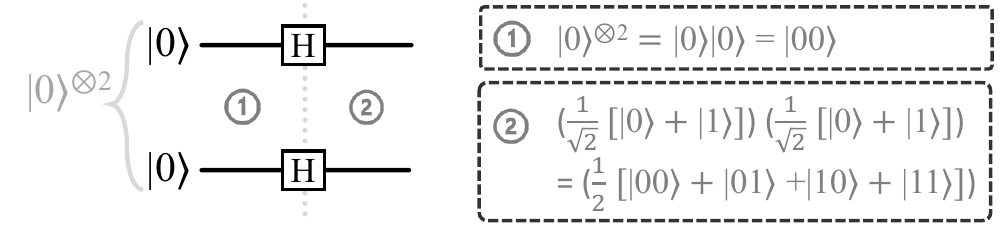}
    \vspace{-8mm}
    \caption{The presented circuit illustrates a fundamental instance of quantum superposition. It commences with an initial 2-qubit state, \raisebox{.5pt}{\textcircled{\raisebox{-.9pt} {\textbf{1}}}}, and culminates in a superposition state, \raisebox{.5pt}{\textcircled{\raisebox{-.9pt} {\textbf{2}}}}, demonstrating the essential properties of quantum systems. \raisebox{.5pt}{\textcircled{\raisebox{-.9pt} {\textbf{1}}}}: $\ket{0}\ket{0} = \ket{00}$; \raisebox{.5pt}{\textcircled{\raisebox{-.9pt} {\textbf{2}}}}: $\frac{1}{2}[\ket{00} + \ket{01} + \ket{10} + \ket{11}]$.}
    \label{fig:superposition}
    \vspace{-5mm}
\end{figure}

\paragraph{\textbf{Quantum Superposition}} Superposition is a phenomenon that allows both the computational basis states $\ket{0}$ and $\ket{1}$ to exist in quantum Hilbert space at the same time. A qubit state can be put into superposition using Hadamard (H) gate. If the initial qubit state is $\ket{0}$ ($\ket{1}$), then the superposition state becomes $\frac{1}{\sqrt{2}}(\ket{0}+\ket{1})$ ($\frac{1}{\sqrt{2}}(\ket{0}-\ket{1})$) after the H gate. We present an example in Fig. \ref{fig:superposition} to generate a superposition of all the basis states for a two-qubit system.

\paragraph{\textbf{Quantum Algorithms and Applications}} The potential of quantum computing has been illustrated by a number of quantum algorithms. Examples include Grover's method for exploring unsorted databases \cite{grover1996fast}, Shor's algorithm for factoring large integers \cite{shor1994algorithms}, and the quantum simulation algorithms \cite{lloyd1996universal}. Among other applications, these algorithms have substantial effects on cryptography, optimization, and quantum system simulation. Combinatorial optimization issues can be resolved using variational quantum-classical algorithms such as, Quantum Approximate Optimization Algorithm (QAOA) \cite{farhi2014quantum}, Variational Quantum Eigensolver (VQE)  \cite{tilly2022variational} and Quantum Machine Learning (QML) models like Quantum Support Vector Machine (QSVM) \cite{rebentrost2014quantum} and quantum principal component analysis (QPCA) \cite{schuld2015introduction}. 

\begin{figure}[t]
    \centering
    \includegraphics[width=1\linewidth]{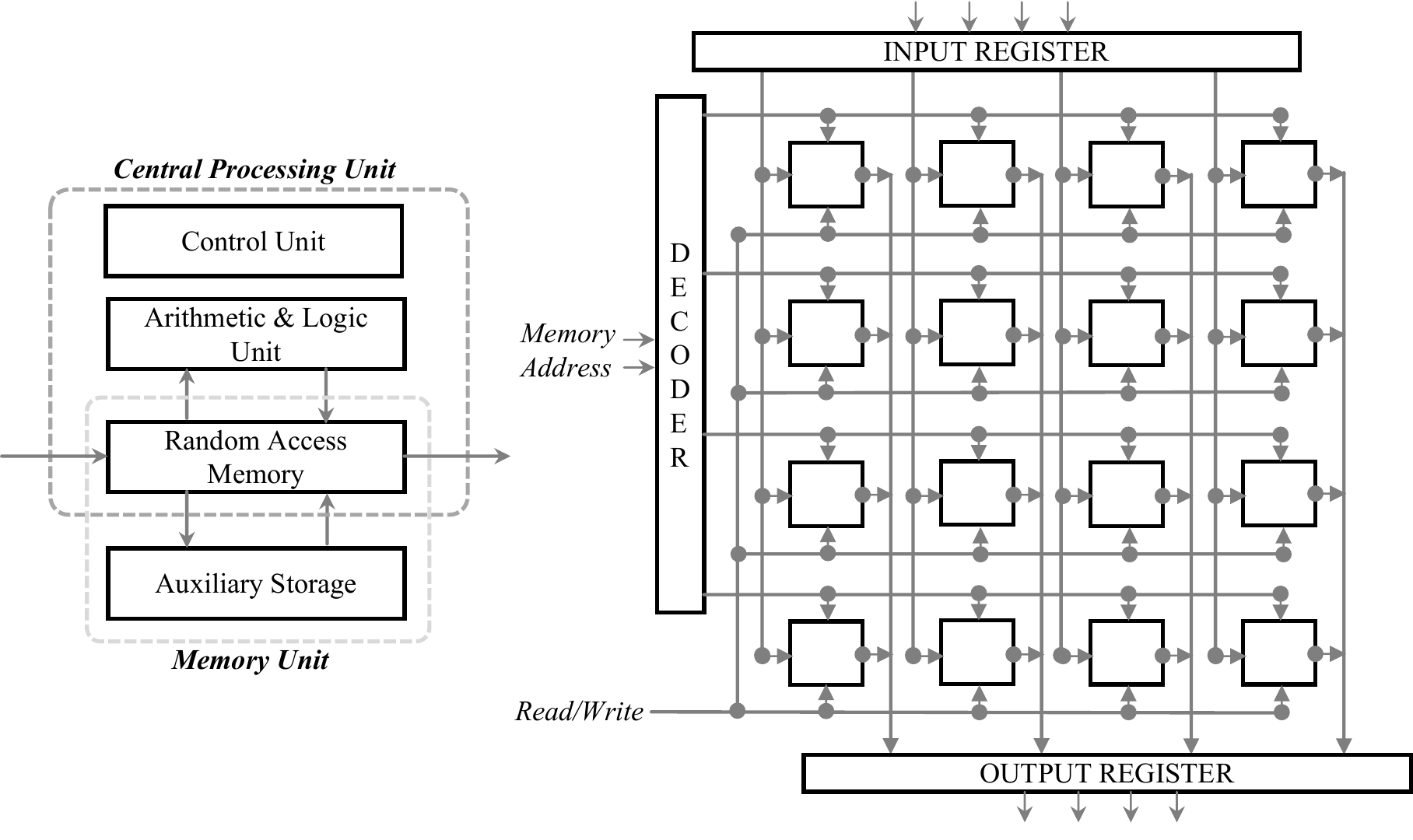}
    \vspace{-7mm}
    \caption{
    \textbf{\textit{Left:}} Depiction of the placement of RAM within the memory hierarchy, highlighting its proximity to the CPU. The RAM's speed can be attributed to this closeness, as it serves as an intermediary between the CPU and auxiliary memory systems. \textbf{\textit{Right:}} A detailed representation of the various functional components within a RAM, illustrating their organization and interconnections. \textit{Memory Array:} It is made up of a grid of rows and columns that stand in for memory cells used to store data. One piece of information is stored in each cell. \textit{Input Register:} During a write operation, it temporarily stores the data that will be stored in the memory array. \textit{Output Register:} During a read operation, it temporarily stores the data that was read from the memory array. \textit{Decoder:} This component takes the memory address from the address bus and converts it into row and column coordinates so that it can access the associated memory cell. To demonstrate the flow of row and column signals, arrows from the decoder should point in the direction of the memory array. \textit{Control Bus:} This transmits read and write enable signals to the memory array to control data access activities.}
    \label{fig:class_ram}
    \vspace{-5mm}
\end{figure}

\subsection{Classical RAM}

Desktop computers can often slow down when running data-intensive applications.
To address this issue, one solution is to install additional Random Access Memory (RAM), which can provide a temporary storage medium for the central processing unit (CPU) to retrieve data quickly in any order while executing a program. It is a volatile 'read/write' memory that stores data temporarily while the computer is operational. When the computer is switched off, the stored data is lost due to its volatile nature. RAM is more efficient than hard drive storage for temporary storage due to faster access time. The fundamental capability of any computing device is the ability to store and manipulate information in a series of memory cells organized in an array \cite{hey1999richard}. RAM is the most well-known architecture for such a memory array since it allows each cell to be addressed \cite{jaeger1997microelectronic}.

A memory array, an input register, and an output register constitute RAM. The memory cells are organized into rows and columns, with each cell holding one bit of data. Data is accessed and manipulated using address lines, data lines, and control lines (read and write enable signals). When the CPU needs to access data from the memory, it sends the memory address through the address lines. Depending on the read or write signal, the data is either retrieved from the memory cell (read operation) or stored in the memory cell (write operation) \cite{patterson2016computer}. The contents of a memory cell are recovered and transferred to the output register when the address of that cell is loaded into the address register (a procedure known as `decoding'). A traditional RAM requires effective data storage, retrieval, and manipulation in order to function. The two main types of RAM, Static Random Access Memory (SRAM), and Dynamic Random Access Memory (DRAM), have unique characteristics that determine their use in different applications \cite{stallings2003computer}. Fig. \ref{fig:class_ram} illustrates the position of RAM within the memory hierarchy and presents a functional block diagram showcasing its key components and their interactions.

\section{Fundamentals of QRAM} \label{qram_fundamentals}

\begin{table}[t]
\caption{A comparison of classical RAM and quantum RAM.}
\begin{tabular}{||c|c|c||}
\hline
\textbf{Attributes}                                                                & \textbf{Classical RAM}                                                      & \textbf{Quantum RAM}                                                                                                    \\ \hline \hline
\textit{Information storage}                                                       & Classical bits ($0/1$)                                                      & \begin{tabular}[c]{@{}c@{}}Qubits\\ ($\ket{\psi}=\alpha\ket{0}+\beta\ket{1}$)\end{tabular} \\ \hline
\textit{\begin{tabular}[c]{@{}c@{}}Access mechanism\\ implementation\end{tabular}} & \begin{tabular}[c]{@{}c@{}}Using transistors \\ and capacitors\end{tabular} & \begin{tabular}[c]{@{}c@{}}Encoding into \\ superposition\end{tabular}                                                  \\ \hline
\textit{Read operation}                                                            & Read signal                                                                 & \begin{tabular}[c]{@{}c@{}}Quantum swap \\ operation\end{tabular}                                         \\ \hline
\textit{Write operation}                                                           & Write signal                                                                & Qubits in input register                                                                                                \\ \hline
\textit{Gate activations}                                                          & $\Theta$($2^n$); $n=\#$bits                                                 & $\Theta$($n$); $n=\#$qubits                                                                                             \\ \hline
\textit{Error correction}                                                          & Repetition codes                                                            & Surface codes                                                                                                           \\ \hline
\textit{Scalability}                                                               & Increasing $\#$bits                                                         & Increasing $\#$qubits                                                                                                   \\ \hline
\end{tabular}
\label{tab:ram_qram_tab}
\vspace{-7mm}
\end{table}

\begin{figure*}[t]
    \centering
    \includegraphics[width=0.89\linewidth]{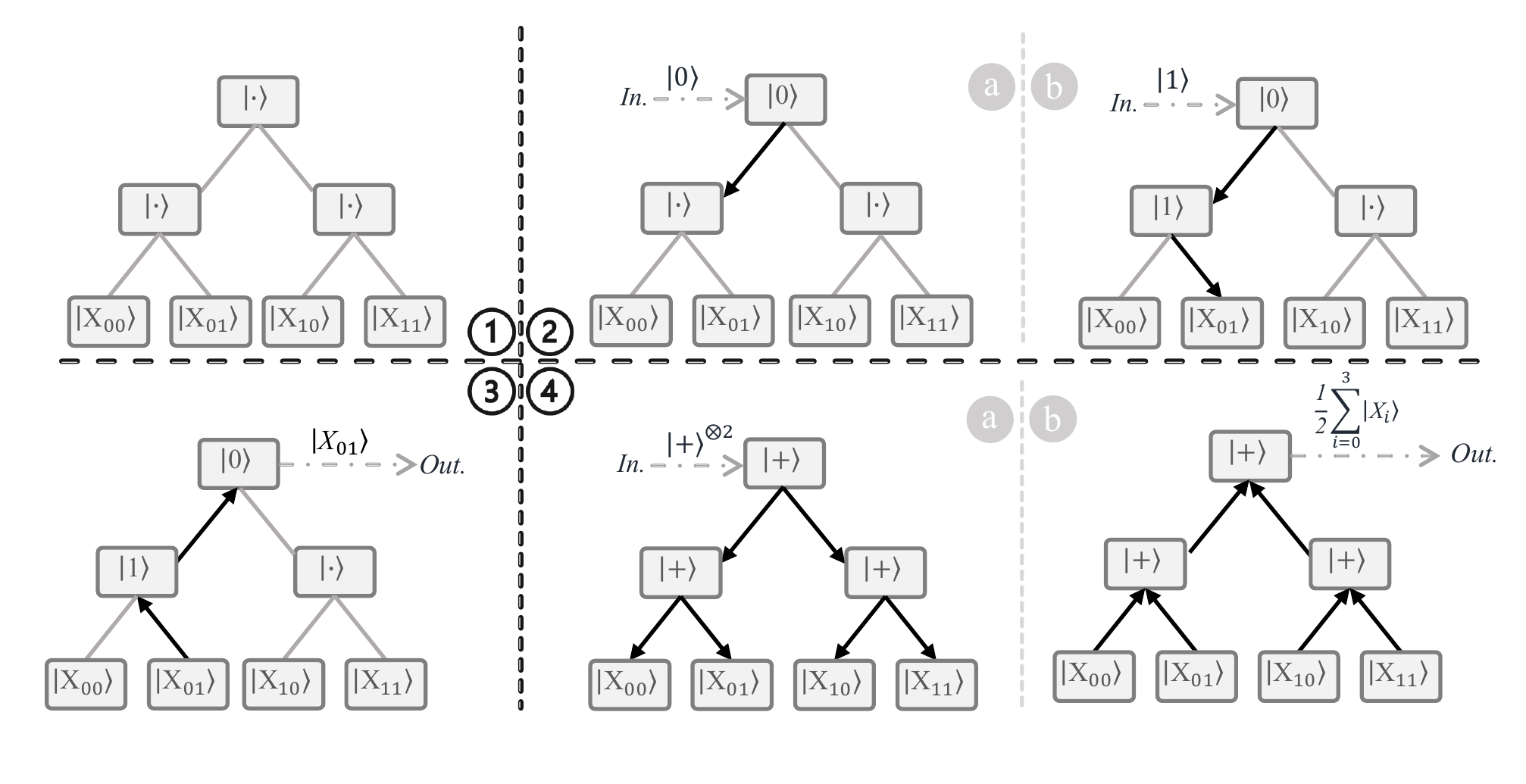}
    \vspace{-10mm}
    \caption{Working of a bucket-brigade QRAM with 2 address lines and 4 memory cells. \raisebox{.5pt}{\textcircled{\raisebox{-.9pt} {\textbf{1}}}} Initial state of the QRAM, all quantum switches are initialized to $\ket{\cdot}$ state which is a waiting state where the quantum switch waits for incoming qubit states of the memory address to be accessed. \raisebox{.5pt}{\textcircled{\raisebox{-.9pt} {\textbf{2}}}} Input register activates switches for output register to access data in address $\ket{01}$ ($\ket{X_{01}}$). The address qubits are sent in a sequential top-down fashion starting from the Most Significant Bit (MSB) all the way to the Least Significant Bit (LSB). In the example shown, first MSB qubit $\ket{0}$ is sent that changes the state of the root quantum switch \raisebox{.5pt}{\textcircled{\raisebox{-0.5pt} {\textbf{a}}}}, followed by the LSB qubit $\ket{1}$ that routes the switch to the direction of the memory cell $\ket{X_{01}}$ \raisebox{.5pt}{\textcircled{\raisebox{-.9pt} {\textbf{b}}}}. \raisebox{.5pt}{\textcircled{\raisebox{-.9pt} {\textbf{3}}}} Output register reads data $\ket{X_{01}}$ via route of activated quantum switches. \raisebox{.5pt}{\textcircled{\raisebox{-.9pt} {\textbf{4}}}} Superposition of all addresses turning on all quantum switches \raisebox{.5pt}{\textcircled{\raisebox{-.5pt} {\textbf{a}}}} to read superposition of all the data \raisebox{.5pt}{\textcircled{\raisebox{-.9pt} {\textbf{b}}}}. Note that $\ket{+}=\frac{1}{\sqrt{2}}(\ket{0})+\ket{1})$., \textit{In.}: Input register, \textit{Out.}: Output register.} 
    \vspace{-7mm}
    \label{fig:buc_bri_qram_working}
\end{figure*}

A QRAM is a memory element analogous to RAM that is able to store data in quantum format. Similar to RAM, a QRAM has three components, the input (or address) register, the output (or data) register, and the memory arrays. The difference here though is that the input and output registers are composed of qubits instead of bits, while the memory arrays can be either classical or quantum depending on the usage of QRAM \cite{giovannetti2008quantum}. For example, for the two fanout QRAM implementations in \cite{giovannetti2008architectures} the optical implementation has 1-bit classical memory cells that change the polarization of the output register photons based on the bit value, while the phase gate implementation uses two superconducting qubits in a single memory cell (one for storing information, one for extracting information). Table \ref{tab:ram_qram_tab} shows the difference between RAM and QRAM. Another key difference in QRAM is the way memory access is performed. Rather than accessing a single memory location at a time, QRAM uses superposition to simultaneously access multiple memory locations. This is made possible by leveraging the power of quantum Hilbert space, where all memory addresses are first loaded into superposition. The overall state is then passed through the QRAM to obtain another superposition state, this time with addresses and data combined. 
Let's say that we have $n$ qubits and consequently $N=2^n$ address lines. All the addresses will be represented as basis states, from $\ket{0}$ to $\ket{N-1}$ \cite{giovannetti2008quantum}, and are stored in address register $r$. Each address $\ket{i}$ will have amplitude $\alpha_i$, so the effective superposition of addresses will be $\sum_{i=0}^{N-1}\alpha_i\ket{i}_r$. This superposition state is then sent to QRAM and the output is another superposition state, which contains both the address state and the data state picked from data register $o$. If $X_i$ is the data in address $i$, then the output state of QRAM is $\sum_{i=0}^{N-1}\alpha_i\ket{i}_r\ket{X_i}_o$. Effectively, the storage of data in QRAM can be summarized through the following equation 
\begin{equation*}
\begin{split}
&\sum_{i=0}^{N-1}\alpha_i\ket{i}_r \xrightarrow{\text{\textit{QRAM}}} \sum_{i=0}^{N-1}\alpha_i\ket{i}_r\ket{X_i}_o\\
\end{split}
\end{equation*}
However, retrieving the data can be challenging due to the no-cloning theorem \cite{wootters1982single}. This is generally taken care of by performing entanglement operations between memory cell qubits and output register qubits using gates such as SWAP gate or CNOT gate.

With the aforementioned statements, readers may become curious and ponder over several crucial aspects of QRAM, namely: (i) the motivation behind the need for QRAM: \textit{Why do we need a QRAM?} (ii) the configuration of QRAM: \textit{What is the structure of a QRAM?} and (iii) the extent of QRAM's utility and its usage: \textit{Where is a QRAM used?} In the following sub-sections, we provide answers to these questions.

\subsection{Why do we need a QRAM?}

In quantum computing, the fundamental building blocks of computation are quantum states, which can represent information as a superposition of basis states. These quantum states are fragile and sensitive to external disturbances, such as environmental noise and decoherence \cite{schlosshauer2019quantum}, which can cause them to rapidly lose coherence and become unusable for computation. Therefore, it is crucial to be able to efficiently store and retrieve quantum states themselves in order to execute quantum algorithms.

Classical memory devices are not suitable for storing quantum states since they will require collapsing of the wavefunction with a measurement operation \cite{von2018mathematical}. The collapse of the wavefunction destroys the superposition of states and causes the quantum state to take on a singular classical value (either 0 or 1), which can be stored in classical RAM but no longer be valuable for quantum computation. 
QRAM is a potential solution to this problem as it allows quantum states to be stored and retrieved efficiently without collapsing the superposition of states. This is accomplished by using quantum mechanics to encode information in a way that is resistant to decoherence and other sources of noise \cite{hann2021resilience}. This allows quantum states to be stored and retrieved with minimal error, making QRAM an essential component of quantum computing technology.

QRAM can also be potentially useful for loading classical data into quantum Hilbert space. Hybrid quantum-classical optimization algorithms in the field of QML often require the conversion of classical data in Euclidean space (e.g., image datasets like MNIST, Iris, CIFAR-10/100, etc.) to quantum Hilbert space into quantum states. This is achieved using encoding methods such as angle embedding, amplitude embedding, and basis embedding \cite{schuld2018supervised}. Amplitude embedding embeds $2^n$ classical features on $n$ qubits, while angle and basis embeddings embed $n$ classical features on $n$ qubits. A problem with these methods, however, is that they are rather simplistic in nature and do not take the complexity of the dataset into account. A QRAM-based loading of data can potentially address the above issue. 

\subsection{What is the structure of a QRAM?}

Various flavors of QRAM architectures have been proposed as described below:

\paragraph{\textbf{Bucket-Brigade QRAM}} The very first proposal of a QRAM \cite{giovannetti2008quantum} implemented a bifurcation graph-based structure as compared to the traditional d-dimensional lattice of memory arrays (shown in Fig. \ref{fig:class_ram}). It is called bucket-brigade QRAM, and the bifurcation graph for this QRAM is a binary tree with the leaf nodes as the memory cells, and the rest of the nodes as switches to route the address state to the correct cell. Overall, there are three main components of this QRAM: the \textit{input register}, the \textit{QRAM itself}, and the \textit{output register}. 
Note, input/index/address register and output/data register/quantum bus are used interchangeably in the literature. For ease of understanding, we use the terms input and output registers in this paper.
There are primarily two cases to explain how a bucket-brigade QRAM works. First, is when there is only a single address in the input register, and second, is when there is a superposition of addresses in the input register. We explain them in the following paragraphs:

\textbf{Single address case:} Let's consider a QRAM that supports two addresses (two qubits) and four memory cells for storage. The bifurcation graph for this QRAM at the start is shown in Fig. \ref{fig:buc_bri_qram_working}, with the quantum switches initialized at the wait state, and the four memory cells present at the leaf nodes. Each quantum switch is a three-level system with states $\ket{\cdot}$, $\ket{0}$, and $\ket{1}$ unlike a qubit which is a two-level system ($\ket{0}$ and $\ket{1}$). This three-level system is often referred to as a qutrit and is inspired by the classical three-level system trit, which are generally tri-state logic multiplexers \cite{horowitz1989art}. The significance of the wait state $\ket{\cdot}$ in each quantum switch is that whenever a qubit state (either $\ket{0}$ or $\ket{1}$) is received by the switch, it changes from $\ket{\cdot}$ to the received state. This helps to achieve that the next time the same switch receives another qubit state, it will route the qubit state to one of its children's node switches. The wait state ensures that un-accessed memory cells are not disturbed. The direction of routing depends on the state of the qubit. Typically $\ket{0}$ ($\ket{1}$) routes the next state to the left (right) child. 

\begin{figure}[t]
    \centering
    \includegraphics[width=\linewidth]{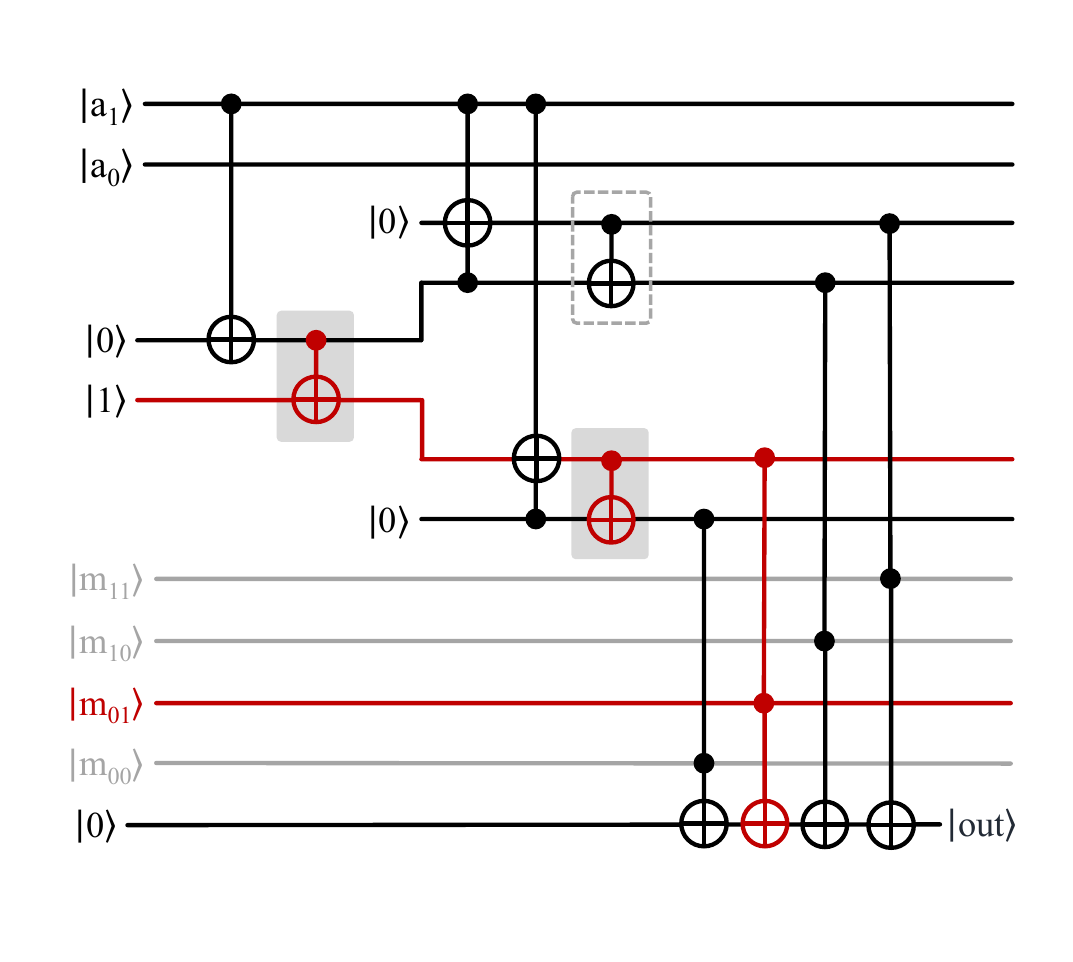}
    \vspace{-15mm}
    \caption{Circuit-based implementation of a bucket-brigade QRAM. Data in memory cell $m_{01}$ with address $\ket{01}$ is being accessed via a series of CNOT and Toffoli gates performing intermediate computation on ancilla qubits. Note that the CNOT gates highlighted in red are the ones getting activated and the red path represents the active route of the QRAM.}
    \label{fig:bucket_brigade_qram_circuit}
    \vspace{-6.5mm}
\end{figure}

Next, we show the example of an incoming address $\ket{01}$ accessing the initialized QRAM. The address state comes from the input register in a sequential fashion from the Most Significant Bit (MSB) to the Least Significant Bit (LSB). Since the address is $\ket{01}$, the MSB is $\ket{0}$ and the LSB is $\ket{1}$. Therefore, state $\ket{0}$ is first sent to the root node switch of the QRAM. Since the root node switch is in the $\ket{\cdot}$, it changes state to $\ket{0}$ (Fig. \ref{fig:buc_bri_qram_working}.2(a)). Next, the LSB state $\ket{1}$ arrives at the root node switch which routes it to the left child. The left child is then activated to state $\ket{1}$ (Fig. \ref{fig:buc_bri_qram_working}.2(b)). In this way, all the address qubits in the input register are used to create a route to memory cell $\ket{X_{01}}$. Along with the graph-based implementation, we also show the circuit-based implementation of the bucket-brigade QRAM with two address lines and four memory cells (inspired from \cite{arunachalam2015robustness}) in Fig. \ref{fig:bucket_brigade_qram_circuit}.

\begin{figure}[t]
    \centering
    \includegraphics[width=\linewidth]{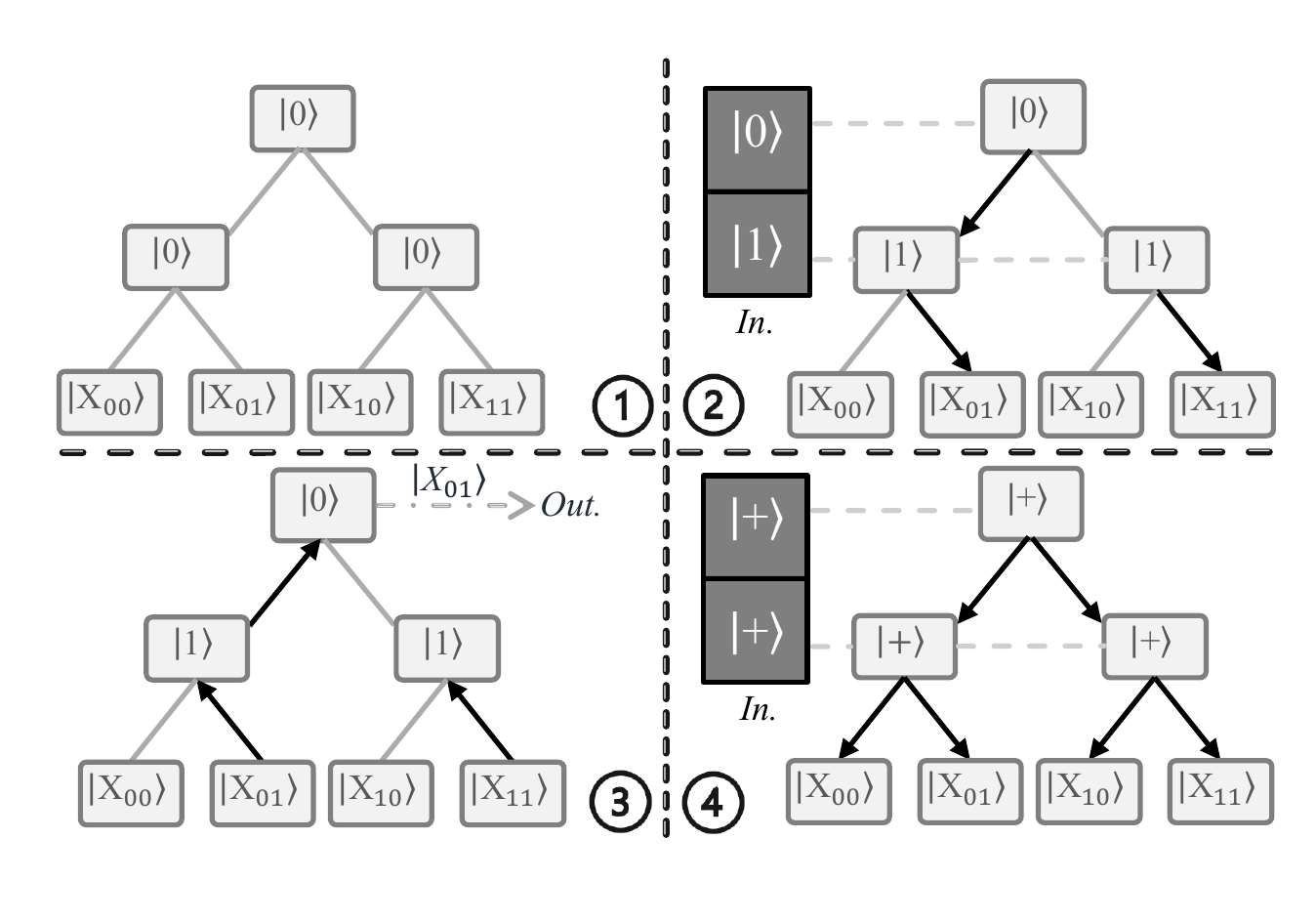}
    \vspace{-13mm}
    \caption{Working of a fanout QRAM. \raisebox{.5pt}{\textcircled{\raisebox{-.9pt} {\textbf{1}}}} Fanout QRAM initialization with all quantum switches initialized to $\ket{0}$ state. \raisebox{.5pt}{\textcircled{\raisebox{-.9pt} {\textbf{2}}}} Address qubits in the input register controlling the state of their respective quantum switches and creating a path from root node switch to the desired memory cell $X_{01}$. \raisebox{.5pt}{\textcircled{\raisebox{-.9pt} {\textbf{3}}}} Data in memory cell $X_{01}$ being accessed in output register via the active route of switches. \raisebox{.5pt}{\textcircled{\raisebox{-.9pt} {\textbf{4}}}} Superposition of addresses switching on all routes to access contents of all memory cells.}
    \label{fig:fanout_qram_working}
    \vspace{-6mm}
\end{figure}

After the creation of the route, the data in the output register can either be stored or read in the routed memory cell. In Fig. \ref{fig:buc_bri_qram_working}.3, we show the example of reading contents of address $\ket{01}$ ($\ket{X_{01}}$) from the memory cell to the output register via the route. Note that if one wants to store new data, the direction of routing will be opposite i.e., from the output register to the memory cell. 

\textbf{Superposition of addresses:} In this case, all the qubits will be present in a superposition state similar to the case shown in Fig. \ref{fig:superposition}. When a qubit in superposition encounters a quantum switch, the switch will also change from $\ket{\cdot}$ to superposition state $\ket{+}=\frac{1}{\sqrt{2}}(\ket{0}+\ket{1})$. Because the superposition state has the presence of both $\ket{0}$ and $\ket{1}$ state, the quantum switch ends up activating both the left and right routes. In this way, when all the superposition address qubits arrive on all the quantum switches, the routes to all memory cells get activated (Fig. \ref{fig:buc_bri_qram_working}.4(a)). Next, when the read operation on the output register is performed, contents from all the memory cells traverse from all the activated routes and load a superposition of all the data $\frac{1}{2}\sum_{i=0}^{3}\ket{X_{i}}$ on the output register. Compared to a classical RAM, the advantage provided by bucket-brigade QRAM is that for $n$ address lines the classical RAM requires O($2^n$) transistor activations for accessing data in a single address, while the QRAM takes only O($n$) quantum switch activations. Furthermore, at a comparable classical cost of O($2^n$) quantum switch activations, the QRAM can read data from all the addresses.  

\paragraph{\textbf{Fanout QRAM}} A follow-up paper \cite{giovannetti2008architectures} on the original bucket-brigade QRAM work \cite{giovannetti2008quantum} presents architectural implementations of bucket-brigade QRAM along with another QRAM termed `fanout' QRAM. Fanout QRAM is taken directly from its classical equivalent fanout RAM, where $k^{th}$ address bit controls $2^k$ switches. Usually, for $n$-bit binary address, the MSB is considered $0^{th}$ address bit and the LSB is considered $n-1^{th}$ address bit. The quantum version of the fanout RAM has $k^{th}$ address qubit controlling $2^k$ quantum switches. A difference between the quantum switches of bucket-brigade QRAM and the fanout QRAM is that while bucket-brigade QRAM requires qutrits, the fanout QRAM requires only a two-level system so qubits are used as the quantum switches. Initially, all the quantum switches are initialized to $\ket{0}$ state. Next, the address qubits present in the input register are used to change the state of the quantum switches. All the quantum switches connected to an address qubit in state $\ket{0}$ will remain at state $\ket{0}$ while the ones connected to an address qubit in state $\ket{1}$ will change their state to $\ket{1}$. 

To explain the functionality of the fanout QRAM, we once again consider a QRAM with two address lines and four memory cells. The bifurcation graph for the QRAM with the quantum switches initialized to $\ket{0}$ state is shown in Fig. \ref{fig:fanout_qram_working}.1. The input register like the previous case can contain either a single address or a superposition of addresses. 

\textbf{Single address case:} Let us first take the simple case of single address access, where the memory cell in address $\ket{01}$ ($X_{01}$) is being addressed. The MSB address state $\ket{0}$ has index $0$ and it will control $2^0=1$ quantum switch, which is the root node switch. The LSB address state $\ket{1}$ has index $1$ so it will control $2^1=2$ quantum switches which are the two child switches of the root node switch. The root node switch will stay at state $\ket{0}$ while the child switches will change state to state $\ket{1}$ (Fig. \ref{fig:fanout_qram_working}.2). As a consequence of this, all the quantum switches are activated, but only a single complete path from the root node to the memory cell ($X_{01}$ in this case) is active. After this, the contents from the memory cell are either updated from or loaded into the output register via the active path (Fig. \ref{fig:fanout_qram_working}.3). 

\textbf{Superposition of addresses:} Extrapolating the above process to a superposition of addresses, the quantum switches will also be switched to superposition. As a result of superposition, each quantum switch will activate routes towards both of its child switches. We show this in Fig. \ref{fig:fanout_qram_working}.4. Finally, for a read operation, contents from all the memory cells will traverse all the active routes and the output register will read a superposition of data (similar to Fig. \ref{fig:buc_bri_qram_working}.4(b)). Compared to the bucket-brigade QRAM, the fanout QRAM activates O($2^n$) switches for both single address access as well as for accessing the superposition of all addresses. 

\begin{table}[t]
\caption{Creation of dataset with rotation angle for FF-QRAM.}
\begin{tabular}{||c|c|c|c|c||}
\hline
\textbf{\begin{tabular}[c]{@{}c@{}}Address\\ (A)\end{tabular}} & \textbf{Data (X)} & \textbf{\begin{tabular}[c]{@{}c@{}}Data\\ Value\end{tabular}} & \textbf{\begin{tabular}[c]{@{}c@{}}Normalized\\ Value ($\bm{X_N}$)\end{tabular}} & \textbf{\begin{tabular}[c]{@{}c@{}}$\bm{\theta=}$\\$\bm{2}$arcsin($\bm{X_N}$)\end{tabular}} \\ \hline \hline
$00$             & $x_{00}$          & $2(10)$                                                       & $2/\sqrt{15}=0.51$                                                          & $1.06$                           \\ \hline
$01$             & $x_{01}$          & $3(11)$                                                       & $3/\sqrt{15}=0.77$                                                          & $1.74$                           \\ \hline
$10$             & $x_{10}$          & $1(01)$                                                       & $1/\sqrt{15}=0.25$                                                          & $0.5$                           \\ \hline
$11$             & $x_{11}$          & $1(01)$                                                       & $1/\sqrt{15}=0.25$                                                          & $0.5$                           \\ \hline
\end{tabular}
\label{tab:ff_qram_data}
\vspace{-4mm}
\end{table}

\paragraph{\textbf{Flip-Flop QRAM}} A more recent quantum circuit-based QRAM implementation called flip-flop QRAM (FF-QRAM) has been proposed in \cite{park2019circuit}. The FF-QRAM stores binary data in superposition one by one, such that the overall circuit has exponential circuit depth and linear width in terms of the number of address lines (or address qubits). Let's assume there are $n$ address lines and the size of each binary data in an address is $m$ bits. Then, the QRAM circuit will have circuit depth O($2^n$) and circuit width O($n+m$). Storing a single data point occurs in three stages: the \textit{flip} stage, the \textit{register} stage, and the \textit{flop} stage. The flip stage is a `compute' stage that is used to make all the data and address qubit states to $\ket{1}$ that is being stored, the register stage consists of a multi-controlled rotation gate that stores the data in a register qubit, and the flop stage is an `un-compute' stage which performs the inverse operation of the compute stage on the address and data qubits.

\begin{figure}[t]
    \centering
    \includegraphics[width=\linewidth]{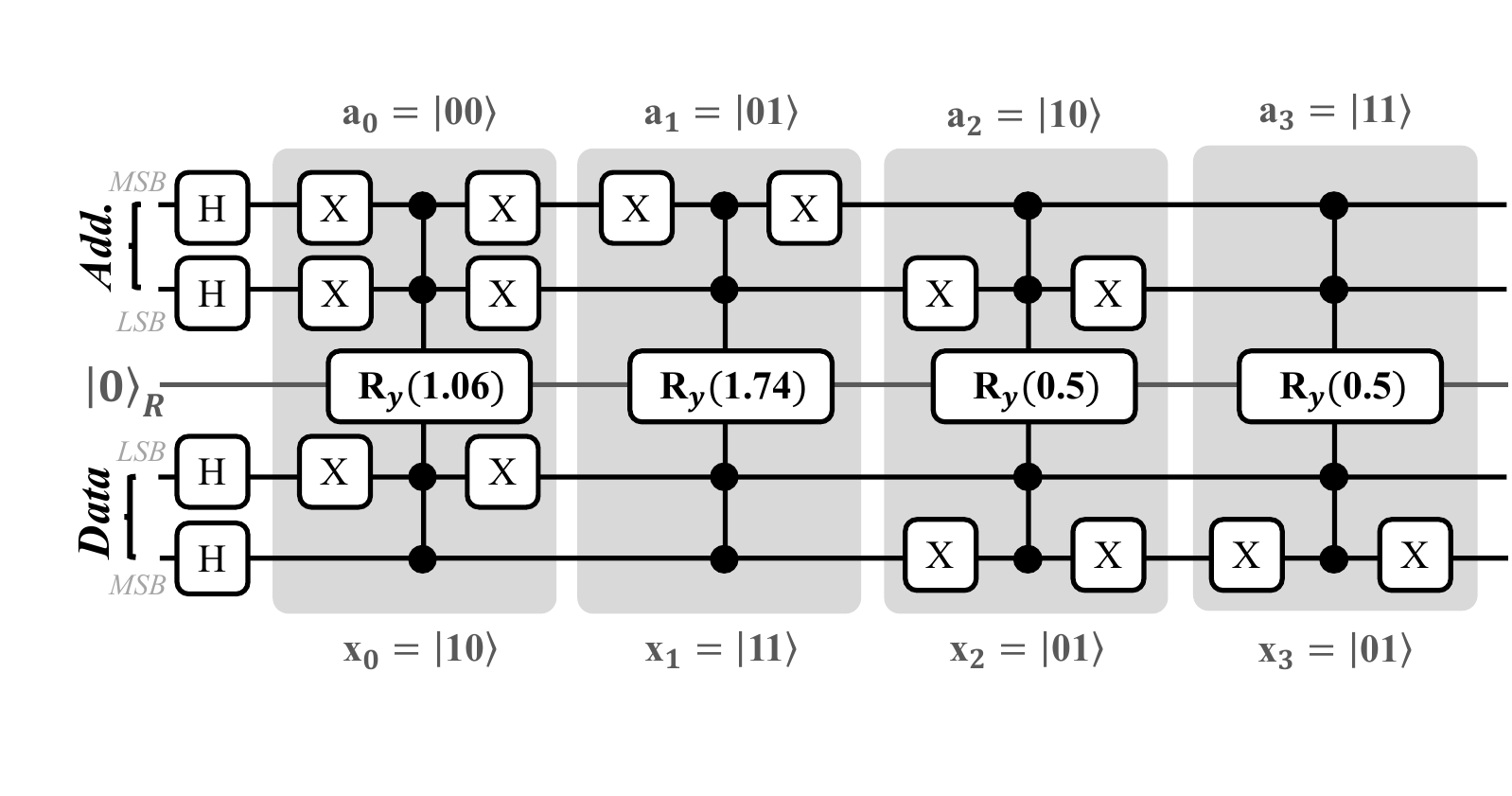}
    \vspace{-14mm}
    \caption{Working of FF-QRAM circuit. The QRAM stores data $10$, $11$, $01$ and $01$ in addresses $00$, $01$, $10$ and $11$ respectively.}
    \label{fig:ff_qram_circuit}
    \vspace{-5mm}
\end{figure}

To explain the working of FF-QRAM, consider a two-address line QRAM with four address-data pairs with each data point of size 2 bits ($n=2$; $m=2$). For each of the four addresses, we have the data as shown in Table \ref{tab:ff_qram_data}, and for each data, we generate its respective rotation angle in two steps, (i) we normalize the data. Given we have the data $\{2,3,1,1\}$, the normalization factor will be $\sqrt{2^2+3^2+1^2+1^2}=\sqrt{15}$, and we divide the dataset with the normalization factor, so new normalized dataset becomes $\frac{1}{\sqrt{15}}\{2,3,1,1\}=\{0.51,0.77,0.25,0.25\}$. (ii) We compute the rotation angle for each data using the normalized values. The rotation angle $\theta_k$ for a datapoint $x_k$ with normalized value $x_{k,n}$ is given as $\theta_k=2$arcsin($x_{k,n}$) \cite{de2020circuit}. For the example shown in Table \ref{tab:ff_qram_data}, the rotation angles will be $2$arcsin($\{0.51,0.77,0.25,0.25\}$)$=\{1.06,1.74,0.5,0.5\}$. 

Once we have the rotation values computed, we can build the FF-QRAM circuit for these address-data pairs. We show the FF-QRAM for our example dataset in Fig. \ref{fig:ff_qram_circuit}. The quantum circuit will have $2$ qubits for address lines, $2$ qubits for data lines, and $1$ qubits for register lines. First, all the address and data qubits are initialized to $\ket{0}$ state and put into superposition using Hadamard gate. After this, the process of storing the data starts. As mentioned earlier, the data is stored one by one in three stages. In the flip stage, which is the compute stage the qubit states of the relevant address and data are flipped to $\ket{1}$ so that the multi-controlled $R_y$ rotation gate is triggered to store the rotation of the desired data. This is achieved using classically controlled-NOT gates \cite{park2019circuit}. Basically, when the classical bit value is $0$ the NOT gate is activated otherwise the NOT gate is not activated. 
We present a simpler version of this gate for ease of understanding as follows: when the classical bit value is $0$, we place an X gate and when the classical value is $1$, we do not place the X gate. Consider the first address-data pair $\ket{00}-\ket{10}$ in Fig. \ref{fig:ff_qram_circuit}. For address qubits, since both the target address lines are in state $\ket{0}$, we put an X gate on both the qubits. Similarly, for the data lines, we put an X gate only on the LSB qubit. Next is the register stage where we add the multi-controlled $R_y$ gate with the computed rotation angle of $1.06$ radians. Finally, we then add the inverse of the flip stage in the flop stage, which will be X gates placed only on those qubits where X gates were placed in the flip stage. The same process is repeated for all the remaining address-data pairs as shown in Fig. \ref{fig:ff_qram_circuit}. The end result of this repetitive process is that the FF-QRAM will have a superposition of addresses and their corresponding data and angle stored in the address, data and register qubits, respectively.

\begin{figure}[t]
    \centering
    \includegraphics[width=0.9\linewidth]{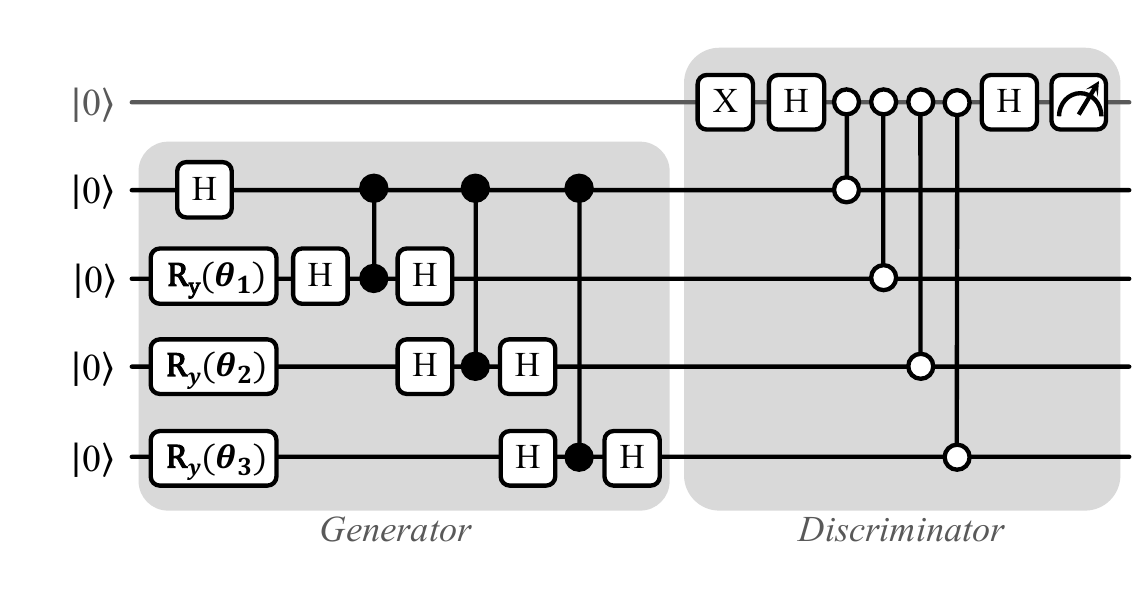}
    \vspace{-3mm}
    \caption{EQGAN variational QRAM circuit for storing superposition of data from class 0. For storing the superposition of data in class 1, the generator ansatz is slightly different.}
    \label{fig:eqgan_qram_circuit}
    \vspace{-4mm}
\end{figure}

\paragraph{\textbf{Entangling Quantum Generative Adversarial Network (EQGAN) QRAM}} EQGAN \cite{niu2022entangling} is pure quantum entanglement-based Generative Adversarial Network (GAN) which is PQC-based and has a quantum generator and a quantum discriminator. They both are trained together with a minimax game. For discriminator model $D$ with parameters $\theta_d$, generator model $G$ with parameters $\theta_g$, real data $\sigma$ and generated data $\rho(\theta_g)$, the minimax problem is given as  
\begin{equation*}
\begin{split}
&\min_{\theta_g}\max_{\theta_d}C(\theta_g,\theta_d)=\min_{\theta_g}\max_{\theta_d}\{1-D_{\sigma}[\theta_d,\rho(\theta_g)]\}\\
\end{split}
\end{equation*}
This EQGAN model is used for variational QRAM as an application, where data points from two Gaussian distributions are stored. The QRAM \cite{niu2022entangling} uses two generators with exponential peak ansatz, one for class 0 and one for class 1 (each class signifies data from one Gaussian distribution), and a swap-test-based discriminator. We show the circuit of this variational EQGAN QRAM in Fig. \ref{fig:eqgan_qram_circuit} which stores data from class 0. For class 1 the PQC is nearly the same, with a slight difference in generator ansatz. Using this generator-discriminator setup, in constant O($1$) gates the QRAM is able to store data into superposition approximately. Another advantage of this QRAM is observed in a classification task. Without QRAM, training the data on a Quantum Neural Network (QNN) yielded an average 45\% classification accuracy, and with the QRAM augmented, the average classification accuracy increased to around 65\%.

\paragraph{\textbf{Qudits-based memory}} Qudits are higher-state quantum units that contain more than two computational basis states. While a qubit in superposition can be represented as $\ket{\psi}=\alpha\ket{0}+\beta\ket{1}$, the superposition in a qudit with $d$ computational basis states is represented as
\begin{equation*}
\begin{split}
&\ket{\psi}=\alpha_0\ket{0}+\alpha_1\ket{1}+....\alpha_{d-1}\ket{d-1}=\sum_{i=0}^{d-1}\alpha_i\ket{i}\\
\end{split}
\end{equation*}
Recent works such as, \cite{gokhale2019asymptotic, baker2020efficient} propose qudits-based quantum memory where qubits are compressed onto qudits temporarily in their higher states using reversible compression circuits. When unused, the qudits can be used as ancilla bits somewhere else. During computing, the qudits can be reverted back to qubits by performing the inverse operation of the compression circuit.  

The authors in \cite{baker2020efficient} define two higher state gates analogous to the X-gate. Let's assume that the input qubit state is $\ket{i}$, then (i)$X_{+t}$ gate performs the operation $\ket{i}\xrightarrow{X_{+t}}\ket{(i+t)\text{ mod } d}$ and (ii) $X_{ij}$ gate performs $\ket{i}\xrightarrow{X_{ij}}\ket{j}$ and $\ket{j}\xrightarrow{X_{ij}}\ket{i}$. There are also the two-qudit controlled versions of these gates where the control qudit has a reference control state and the target qudit has the gate. They also introduce an x-y-z qudit-qubit compression scheme that is dependent on the radix of the input and output qudits. Here, $x$ is the radix of the input qudits, $y$ is the radix of the output qudits and $z$ are the number of ancilla generated. The compression scheme should follow $x^a\leq y^b$, such that $0<b<a$ and $a-b=z$, where $a$,$b$ are integers. $x^a$ denotes the number of computational basis states of the input and $y^b$ denotes the number of computational basis states of the output. A natural restriction is that the computational basis states of the output should be higher than that of the input. Another restriction is that the number of input qudits $a$ should be greater than the number of output qudits $b$ for enabling compression. As a result of this compression, a total of $a-b=z$ ancilla qubits are generated.

\begin{figure}[t]
    \centering
    \includegraphics[width=\linewidth]{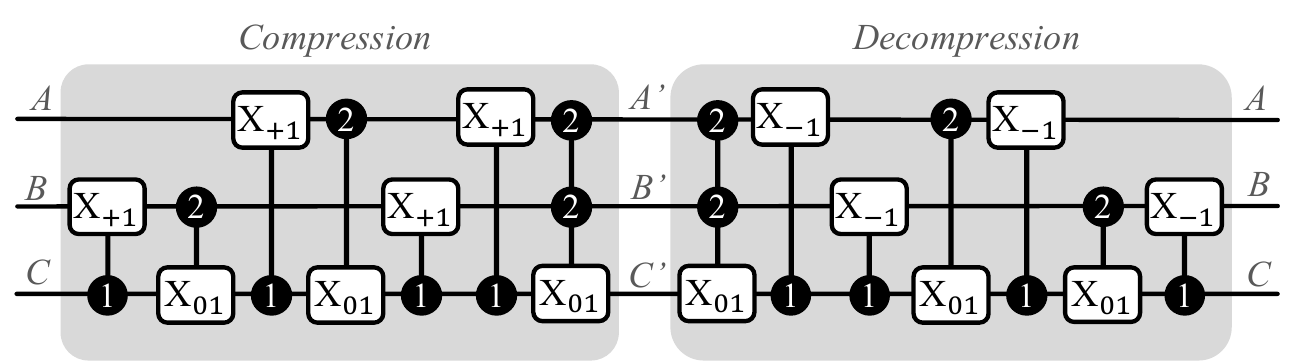}
    \vspace{-7mm}
    \caption{Compression and decompression circuits for a $2$-$3$-$1$ compression scheme \cite{baker2020efficient}. The compression circuit compresses the contents of three qubits (A,B,C) in two qutrits (A',B') and generates a free ancilla in state $\ket{0}$ (C'). The decompression circuit again reverts the qutrits and ancilla back to the original three qubits.}
    \label{fig:qdit_circuit}
\end{figure}

\begin{table}[t]
\centering
\caption{Truth table for $2$-$3$-$1$ compression scheme \cite{baker2020efficient}.}
\begin{tabular}{||c|c|c||c|c|c||}
\hline
\textbf{A} & \textbf{B} & \textbf{C} & \textbf{A'} & \textbf{B'} & \textbf{C'} \\ \hline \hline
0          & 0          & 0          & 0           & 0           & 0           \\ \hline
0          & 0          & 1          & 2           & 2           & 0           \\ \hline
0          & 1          & 0          & 0           & 1           & 0           \\ \hline
0          & 1          & 1          & 0           & 2           & 0           \\ \hline
1          & 0          & 0          & 1           & 0           & 0           \\ \hline
1          & 0          & 1          & 2           & 1           & 0           \\ \hline
1          & 1          & 0          & 1           & 1           & 0           \\ \hline
1          & 1          & 1          & 1           & 2           & 0           \\ \hline
\end{tabular}
\label{tab:231_comp_truth_table}
\vspace{-5mm}
\end{table}

A simple example is the conversion of qubits ($d=2$) to qutrits ($d=3$). Three qubits can store $2^3=8$ computational basis states, and two qutrits can store $3^2=9$ computational basis states. So three qubits can be compressed into two qutrits and the leftover qubit can be generated as an ancilla that can be used in other circuits. For this example, we have $x=2$, $y=3$, and $z=1$, so it is a $2$-$3$-$1$ compression scheme. We show the compression and decompression circuits of this scheme in Fig. \ref{fig:qdit_circuit}. The compression circuit consists of controlled $X_{+1}$ and $X_{01}$ conditioned either on $\ket{1}$ or $\ket{2}$ states. The decompression circuit has the gates of the compression circuit in reverse order with an added difference of having controlled $X_{-1}$ gates instead of controlled $X_{+1}$ gates. The truth table of the $2$-$3$-$1$ compression scheme is shown in Table \ref{tab:231_comp_truth_table}. By substituting values of the qubits A, B, C in the circuit and performing higher level qudit operations, it can be verified that the compression yields corresponding A', B' (qutrits) and C' (ancilla in $\ket{0}$ state) entries from the truth table and decompression yields back original values of A, B, and C.

\begin{figure}[t]
    \centering
    \includegraphics[width=0.75\linewidth]{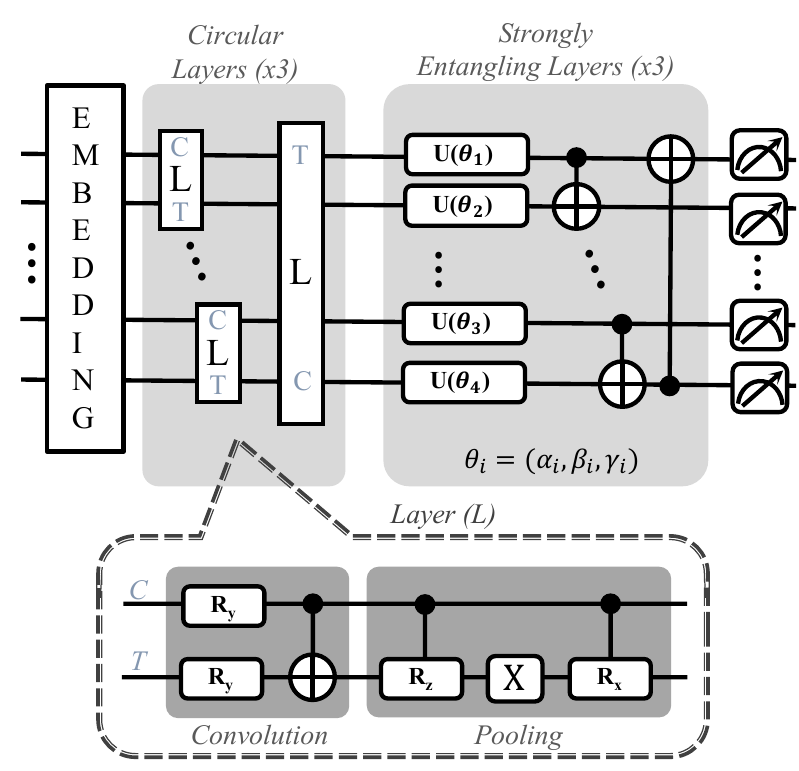}
    \vspace{-3.5mm}
    \caption{PQC structure of approximate PQC-based QRAM. In the given context, `C' symbolizes the control qubit, while `T' stands for the target qubit.}
    \label{fig:pqc_qram_circuit}
    \vspace{-5mm}
\end{figure}

\paragraph{\textbf{Approximate PQC-based QRAM}} A trainable PQC-based QRAM \cite{phalak2022approximate} similar to EQGAN QRAM is proposed that is able to store data in the quantum Hilbert space by training the PQC. Compared to the EQGAN, the approximate PQC-based QRAM does not store data in superposition but one-by-one in sequential order and is able to store more complex datasets such as image datasets like the UCI digits dataset. The approximate PQC-based QRAM is also used for the pure storage of binary data. The detailed PQC of the approximate QRAM is shown in Fig. \ref{fig:pqc_qram_circuit}. It consists of an embedding scheme such as angle, amplitude, or basis embedding to load classical data, followed by 3 sets of circular layers and strongly entangling layers respectively. 
It has been noted that loading images from QRAM and sending them to a QNN yields faster convergence of classification (by $6^{th}$ epoch) as compared to loading images without QRAM (around $15^{th}$ epoch), and for pure storage the QRAM is able to store 4-bit binary data without any errors.

\subsection{Where is a QRAM used?}
A QRAM that is able to store and load data in superposition is very helpful for certain classes of quantum algorithms.
\begin{itemize}
    \item \textit{Database search:} Grover's algorithm \cite{grover1996fast} and its more generalized version Quantum Amplitude Amplification and Estimation (QAE) \cite{brassard2002quantum} have been proposed to perform a database search for an element out of $n$ elements with complexity O($\sqrt{n}$). They take data in superposition as input and perform the amplification operation O($\sqrt{n}$) times prior to performing estimation with a reduced number of measurements. 
    \item \textit{Element distinctness:} 
    Given a set of $n$ elements, the element distinctness problem asks whether all $n$ elements in the set are distinct. Classically it takes O($n$ log($n$)) time while quantum algorithms such as \cite{ambainis2007quantum} solve it in O($n^{\frac{2}{3}}$) time.
    \item \textit{Collision detection:} Collision detection is an important problem in cryptography. 
    Given a collision function $H$, the collision detection problem asks to find two distinct inputs $x$ and $y$ such that $H(x) = H(y)$. Quantum versions of the collision detection problem such as \cite{brassard1997quantum} report O($n^{\frac{1}{3}}$) runtime where $n$ denotes the cardinality of the domain of collision function.   
    \item \textit{NAND tree evaluation:} In this problem, a Boolean expression is solved using a tree of NAND gates. For an input of size $n$, quantum algorithms such as \cite{childs2007every} propose a runtime of O($\sqrt{n}$).
    \item \textit{Quantum forking:} In classical operating systems, forking is the process of creating a child process from a parent process which is a copy of it, while retaining the parent process. Quantum forking is a similar idea, where the QRAM output superposition state is forked on ancilla qubits and then both the original state and forked state are multiplied with the same or different unknown unitaries. The new states then undergo a swap-test procedure to verify if the unitaries applied are the same or different \cite{park2019circuit, park2019parallel}.
    \item \textit{Storage of classical data:} As mentioned previously, works such as \cite{niu2022entangling, phalak2022approximate} use a PQC-based QRAM circuit to store classical data such as data from a normal distribution, images, and binary data into quantum Hilbert space by training the PQC like a machine learning model.
\end{itemize}


\begin{figure}[t]
    \centering
    \includegraphics[width=0.8\linewidth]{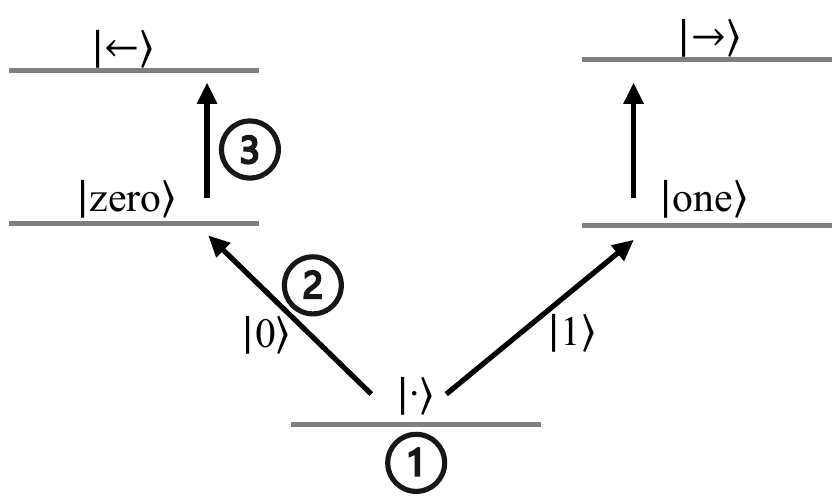}
    \vspace{-4mm}
    \caption{Energy levels of trapped atom-based qutrit switches in bucket-brigade QRAM. \raisebox{.5pt}{\textcircled{\raisebox{-.9pt} {\textbf{1}}}} Initialized state of qutrit. \raisebox{.5pt}{\textcircled{\raisebox{-.9pt} {\textbf{2}}}} First incoming photon getting absorbed and changing the state of the qutrit and routing it either in $\ket{\text{zero}}$ or $\ket{\text{one}}$ direction based on the state of the qubit encoded in the photon. \raisebox{.5pt}{\textcircled{\raisebox{-.9pt} {\textbf{3}}}} Subsequent photons getting absorbed to $\ket{\leftarrow}$ or $\ket{\rightarrow}$ and remitted to the next qutrit based on the state of the previous qubit.}
    \label{fig:bb_energy}
    \vspace{-5mm}
\end{figure}


\section{Practicality of QRAM} \label{qram_application}

Follow-up paper \cite{giovannetti2008architectures} on \cite{giovannetti2008quantum} provide possible physical implementations of bucket-brigade QRAM and fanout QRAM. We first explain these implementations followed by implementation details on the FF-QRAM, qudits-based storage, and trainable PQC-based QRAM. We also show a detailed tabular comparison of different QRAMs in Table \ref{tab:qram_comparison}.

\paragraph{\textbf{Bucket-Brigade QRAM implementation}} For physically implementing the bucket-brigade QRAM, the authors in \cite{giovannetti2008architectures} incorporate (i) address qubits in the input register as photons that can be sent sequentially and (ii) the qutrits as trapped atoms inside cavities. The qubits encoded in the photons traverse the cavity by encountering the trapped atom-based qutrits. The three states of the qutrits are realized as three different energy levels, with the $\ket{\cdot}$ being the lowest energy state, with two higher energy levels $\ket{\text{zero}}$ that is coupled with the left spatial path i.e., to the left qutrit along the bifurcation graph and $\ket{\text{one}}$ that is coupled to the right spatial path to the right qutrit. This coupling is represented using further higher energy states $\ket{\leftarrow}$ (for left spatial path) and $\ket{\rightarrow}$ (for right spatial path). We show the energy diagram of the qutrit switches in Fig. \ref{fig:bb_energy}.

\begin{table*}[t]
\caption{Table of comparison between different QRAM technologies.}
\begin{tabular}{||l|c|c|c|c|c||}
\hline
\textbf{\textit{\textbf{Feature}}}                                                        & \textbf{\textbf{\begin{tabular}[c]{@{}c@{}}Bucket-Brigade\\ QRAM \cite{giovannetti2008quantum}\end{tabular}}}       & \textbf{\textbf{\begin{tabular}[c]{@{}c@{}}Fanout QRAM \cite{giovannetti2008architectures}\end{tabular}}}                                          & \textbf{\begin{tabular}[c]{@{}c@{}}Flip-Flop QRAM \cite{park2019circuit}\end{tabular}}                      & \textbf{\textbf{\begin{tabular}[c]{@{}c@{}}Qudits-based memory \\ \cite{baker2020efficient}\end{tabular}}}                                                                                & \textbf{\begin{tabular}[c]{@{}c@{}}Approximate PQC-based \cite{phalak2022approximate}\\ \&  EQGAN QRAM \cite{niu2022entangling}\end{tabular}} \\ \hline \hline
\textit{Structure}                                                                        & Bifurcation graph                                                                     & Bifurcation graph                                                                                              & Quantum circuit                                                                      & Higher states                                                                                               & \begin{tabular}[c]{@{}c@{}}Parametric Quantum\\ Circuit\end{tabular}                         \\ \hline
\textit{\begin{tabular}[c]{@{}l@{}}Circuit width\\ ($n=\#$address \\ lines)\end{tabular}} & O($2^n$)                                                                              & O($2^n$)                                                                                                       & O($n$)                                                                               & \begin{tabular}[c]{@{}c@{}}Dependent on\\ $d$ (\# qudit states)\end{tabular}                                & O($n$)                                                                                       \\ \hline
\textit{\begin{tabular}[c]{@{}l@{}}Circuit depth\\ ($n=\#$address\\ lines)\end{tabular}}  & O($2^n$)                                                                              & O($2^n$)                                                                                                       & O($2^n$)                                                                             & \begin{tabular}[c]{@{}c@{}}Dependent on \\ $d$ (\# qudit state)\end{tabular}                                & O($1$)                                                                                       \\ \hline
\textit{Unique qualities}                                                                 & \begin{tabular}[c]{@{}c@{}}Qubits are routed\\ in a sequential\\ fashion\end{tabular} & \begin{tabular}[c]{@{}c@{}}Qubits controlling \\ exponential quantum\\ switches\end{tabular}                   & Quantum circuit-based                                                                & \begin{tabular}[c]{@{}c@{}}Reduces requirements\\ of ancilla qubits to 0\end{tabular}                       & \begin{tabular}[c]{@{}c@{}}Trainable like a \\ machine learning \\ model\end{tabular}        \\ \hline
\textit{\begin{tabular}[c]{@{}l@{}}Implementation\\ technology\end{tabular}}              & \begin{tabular}[c]{@{}c@{}}Photons, trapped \\ atoms\end{tabular}                     & \begin{tabular}[c]{@{}c@{}}Photons, microwave\\ cavities\end{tabular}                                          & \begin{tabular}[c]{@{}c@{}}Superconducting qubits,\\ trapped ion qubits\end{tabular} & \begin{tabular}[c]{@{}c@{}}Superconducting qudits,\\ trapped ion qudits, OAM\\ photonic qudits\end{tabular} & \begin{tabular}[c]{@{}c@{}}Superconducting qubits,\\ trapped ion qubits\end{tabular}         \\ \hline
\textit{Drawbacks}                                                                        & \begin{tabular}[c]{@{}c@{}}Exponential circuit\\ width and depth\end{tabular}         & \begin{tabular}[c]{@{}c@{}}Exponential circuit\\ width and depth,\\ susceptible to \\ decoherence\end{tabular} & \begin{tabular}[c]{@{}c@{}}Exponential circuit\\ depth\end{tabular}                  & Unstable higher states                                                                                      & \begin{tabular}[c]{@{}c@{}}Performance degradation under \\ noise (approx. QRAM), store \\ only simple dataset (EQGAN)\end{tabular}                \\ \hline
\end{tabular}
\label{tab:qram_comparison}
\vspace{-4mm}
\end{table*}

Initially, all the qutrits are initialized to the lowest energy state $\ket{\cdot}$. When the first photon traverses through the cavity and reaches the root node switch, it gets absorbed into the higher energy state of the qutrit, either $\ket{\text{zero}}$ or $\ket{\text{one}}$ thereby changing the state of the qutrit depending on the quantum state encoded in the photon. This process is often referred to as the Raman transition, where a photon is scattered by a molecule, resulting in a change in the energy of the photon and the vibrational state of the molecule \cite{feng2002quantum}. This is achieved with the help of strong laser fields \cite{moy1997atom} that help in changing the state of the qutrit from $\ket{\cdot}$ to $\ket{\text{zero}}$ if the photon state is $\ket{0}$ and from $\ket{\cdot}$ to $\ket{\text{one}}$ if the photon state is $\ket{1}$. After this, when the second photon arrives at the root node switch it once again gets absorbed and undergoes a Raman transition, but this time either from $\ket{\text{zero}}$ to $\ket{\leftarrow}$ or $\ket{\text{one}}$ to $\ket{\rightarrow}$ and will be remitted to the qutrit along the correct spatial path based on the state of the qutrit ($\ket{\text{zero}}$: left path, $\ket{\text{one}}$: right path). In this way, all the photons of the input register set the qutrit switches one by one until a path from the root node switch to the desired memory cell is created. The output register then either loads contents from the memory cell or stores new data in it via the created path of qutrit switches. Once the load/store operation is complete, all the qutrits starting from the last node to the root node undergo a final Raman transition sequentially to go back to $\ket{\cdot}$ state.

A recent work \cite{arunachalam2015robustness} proposed a quantum circuit-based implementation of the bucket-brigade QRAM. For $n$ address lines, the quantum circuit requires O($n$) qubits for address, O($2^n$) ancilla qubits for incorporating the quantum switches, O($2^n$) qubits for memory cells and one qubit for readout of the memory cell. We show an implementation of quantum circuit-based bucket-brigade QRAM for two address lines and four memory cells in Fig. \ref{fig:bucket_brigade_qram_circuit} where the memory cell in address $\ket{01}$ is being accessed. First, $\ket{a_1}=\ket{0}$ changes the state of the first ancilla qubit which then changes the state of the next ancilla qubit. Based on the output, the path is then routed to the left child switch, where $\ket{a_0}=\ket{1}$ is used to switch the right ancilla qubit to $\ket{1}$ state. Finally, a set of Toffoli gates with ancilla qubit as one control and the memory cell qubit as another control are used to perform readout. Depending on the address state, only the Toffoli gate controlled by its corresponding memory cell will be triggered. In this case, the Toffoli gate corresponding memory cell $\ket{m_{01}}$ is triggered to perform a readout operation on the readout qubit.

\paragraph{\textbf{Fanout QRAM implementation}} Two implementations, namely optical implementation and phase gate implementation have been proposed for the fanout QRAM \cite{giovannetti2008architectures}. Understanding the implementations in the original paper may be challenging, as it assumes knowledge of optical and cavity-based quantum systems.

In the phase gate implementation, (i) the address qubits in the input register are photons, and (ii) the quantum switches are also photonic qubits trapped inside microwave cavities. Overall, for $n$ address qubits there are O($2^n$) microwave cavities with each cavity containing a photonic qubit. As mentioned earlier, the $k^{th}$ index qubit fans out and controls $2^k$ quantum switches. This is achieved using conditional phase shifters. The MSB address qubit will only polarize the root node photon inside the microwave cavity via the conditional phase shifter attached to it. The next address qubit will polarize two child photons via a conditional phase shifter. This continues until all the photons inside the microwave cavity are polarized. As a result of this, a resonant path will be created from the root cavity to the desired memory cell. Each memory cell consists of two superconducting qubits, such that one is for storing information, and one is for extracting information. The contents of the memory cell are then transferred back to the output register using a SWAP gate via an outgoing photon from the memory cell with the help of the extraction qubit. 

Next, we discuss the optical implementation. Here, (i) the address qubits are atoms trapped in a magneto-optical trap and (ii) the quantum switches are photonic qubits that hit the trapped atomic address qubits one by one. When the first quantum switch photon hits the first address qubit inside the trapped atom, the trapped atom acts as a controller for changing the polarization state of the photon. This photon then passes through a polarization beam splitter and a half-wave plate to transfer this information to another spatial degree of freedom and create two spatial modes. The two spatial modes will be two photonic quantum switches for the next address qubit. Once again, each spatial mode will transfer the state of the address qubit via a change of polarization and create two new modes. This continues until $2^n$ spatial modes are created, one for each memory cell. Out of these, only one spatial mode will be active depending on the address, and the contents of the desired memory cell corresponding to the active spatial mode are swapped out with the contents of the output register using a SWAP gate.

\paragraph{\textbf{EQGAN QRAM implementation}} Since EQGAN QRAM is a quantum circuit-based QRAM, it can be implemented on superconducting and trapped ion qubits. The authors in \cite{niu2022entangling} implement the EQGAN QRAM on 5 qubits of Google's Sycamore superconducting processor such that the readout qubit (top qubit in Fig. \ref{fig:eqgan_qram_circuit}) is the center physical qubit and the rest of the qubits are physically coupled with the readout qubit in the shape of a (+) sign on a grid of qubits.  

\paragraph{\textbf{Qudit implementation}} Qudits are implementable on physical quantum systems that have an infinite spectrum of states, like superconducting qubits (magnetic flux spectrum) \cite{liu2017transferring}, trapped ion qubits (energy band spectrum) \cite{low2020practical}, and Orbital Angular Momentum (OAM spectrum) based photonic qubits \cite{bent2015experimental}. For example, in bucket-brigade QRAM, the qutrit switches are implemented using trapped atoms in a cavity, with the $\ket{\cdot}$ state at a lower energy level and $\ket{\text{zero}}$ and $\ket{\text{one}}$ states at a higher energy level.   

\paragraph{\textbf{Approximate PQC-based QRAM \& Flip-Flop QRAM implementations}} Similar to the EQGAN QRAM, since both of these QRAMs are quantum-circuit based they can be implemented on superconducting and trapped ion qubits. With the quantum circuit known, the QRAM architectures can be replicated on known quantum computing platforms such as Qiskit \cite{ibm_quantum} from IBM, Pennylane \cite{bergholm2018pennylane} from Xanadu, IonQ \cite{ionq} and many more. The users can replicate the quantum circuit and send it either for simulation on a noiseless/noisy simulator (better for approximate PQC-based QRAM \cite{phalak2022approximate} as it is iterative), or run it on actual quantum hardware (better for FF-QRAM \cite{park2019circuit} as it is non-iterative). 
\section{Challenges and Future Direction} \label{qram_future}

In this section, we examine the current limitations and future directions of various QRAM architectures and the challenges. 
Some of the common challenges include:

\begin{itemize}
    \item \textit{Scalability:} Scalability is a major challenge in QRAM designs due to constraints in qubit interactions, quantum memory, and coherence. Increasing memory elements in bucket-brigade, fanout, and FF-QRAMs leads to exponential growth in circuit width and depth. Thus, scalability remains a significant hurdle for large-scale QRAM implementations.
    \item \textit{Noise resilience:} Noise resilience is a crucial challenge in QRAM architectures, as quantum systems are sensitive to environmental noise. In various QRAM types, increasing memory elements results in higher circuit depth and qubit count, making the system more vulnerable to noise. Bucket brigade QRAM is comparatively more noise-resilient than fanout QRAM \cite{hann2021resilience}, while FF-QRAM is susceptible to noise with increasing address lines. PQC-based QRAMs have constant circuit depth but are still prone to noise-related errors, affecting performance on real hardware versus simulation.
    \item \textit{No-cloning theorem:} The no-cloning theorem \cite{wootters1982single, dieks1982communication}, a fundamental quantum mechanics principle, prohibits exact copying of unknown quantum states and poses challenges for various QRAM designs. In bucket brigade and fanout QRAM, the theorem limits duplication of quantum states during memory readout. Solutions like CNOT or SWAP gates are available, but the no-cloning theorem still complicates error correction and redundancy schemes in most QRAM designs, presenting a significant challenge \cite{hann2021practicality}.
    \item \textit{Instability of qudits:} Qudit instability primarily affects qudit-based quantum memory, where qudits are quantum systems with $d > 2$ levels. While qudit-based memory can store more information than qubit-based systems, higher qudit states are unstable and prone to errors \cite{grassl2018quantum, lanyon2008manipulating}. 
    For example, the energy gap between higher states in superconducting qubits is less \cite{krantz2019quantum}. 
    This issue is not directly relevant to qubit-based QRAM designs, but incorporating qudits for increased storage would introduce similar challenges related to qudit instability. 
    \item \textit{Limited applicability:} Some QRAM architectures face limitations due to their novelty, experimental difficulties, or specific focus. For example, FF-QRAM has limited applicability beyond quantum forking due to its targeted design. Similarly, qudits face challenges arising from limited research and increased complexity compared to qubits. Addressing these challenges is essential for advancing broader applications in quantum computing.
\end{itemize}

While current QRAM designs face the above challenges, ongoing research strives to overcome them. Several recent works have made significant progress in the implementation of bucket brigade QRAM. In one study  \cite{casares2020circuit}, researchers construct a circuit implementation of the aforementioned QRAM, demonstrating that when used with classical data, it can quickly and repeatedly prepare arbitrary quantum states once the data is already present in memory. Another work \cite{paler2020parallelizing} discusses the parallelization of queries in a bucket-brigade QRAM, showing that the parallelization method is compatible with surface code quantum error correction. In theory, fault-tolerant bucket-brigade QRAM queries can be performed at speeds comparable to classical RAM. A separate article \cite{arunachalam2015robustness} addresses the robustness of bucket brigade QRAM, revealing that when quantum error correction is applied to the bucket brigade QRAM circuit, the circuit loses its advantage of having a small number of active gates, as error correction operates on all of its components.

The precise evaluation of the hardware expenditure associated with QRAM designs, especially in the realm of fault-tolerant systems, may constitute a significant subject of forthcoming research \cite{hann2021practicality}. It is reasonable to anticipate that when compared to conventional surface code implementations \cite{dennis2002topological}, the hardware expenses and intricacy will be substantially reduced owing to the noise resilience inherent in bucket-brigade QRAM and the implementation of low-overhead fault tolerance techniques utilizing qubits. While there have been notable advancements in hardware efficiency, it remains a challenge to develop a QRAM capable of addressing millions or billions of individual memory elements in the near future. Exploring applications where smaller QRAMs can provide value and conducting tailored resource estimations for these use cases could be key for future progress and development.
\section{Conclusion} \label{qram_conclusion}

Quantum Random Access Memory (QRAM) 
serves as a specialized form of memory that enables direct access and manipulation of quantum states, thereby facilitating expedited and efficient data retrieval and storage within quantum systems. Unlike conventional RAM structures that store information in classical bits, which are incompatible with quantum systems, QRAM operates on the principles of quantum computing. This enables QRAM to store and manipulate quantum data effectively, resulting in considerable acceleration of known quantum algorithms. 
This review provides a thorough assessment of QRAM, emphasizing its importance and practicality within the context of contemporary quantum computing. We explain the fundamentals of quantum computing and conventional RAM, before delving into the foundations of QRAM. 
Five notable types of QRAM designs are outlined, including bucket-brigade QRAM, fanout QRAM, flip-flop QRAM, qudits-based quantum memory, and approximate PQC-based QRAM. By comparing these diverse architectural approaches and carefully analyzing their implementation, the feasibility of QRAM is thoroughly explored. The analysis concludes by discussing the primary challenges associated with QRAM development and future directions.

\bibliographystyle{IEEEtran}
\bibliography{references}

\end{document}